\documentclass[10pt]{article}

\usepackage[english]{babel}
\usepackage[a4paper,top=2cm,bottom=2cm,left=3cm,right=3cm,marginparwidth=1.75cm]{geometry}
\usepackage[backend=biber,style=nature]{biblatex}
\usepackage{amsmath, amssymb, amsfonts, amsthm}
\usepackage{mathrsfs}
\usepackage{cancel}
\usepackage{xfrac}
\usepackage{graphicx}
\usepackage{caption}
\usepackage{subcaption}
\usepackage[colorlinks=true, allcolors=blue]{hyperref}
\usepackage[auth-sc, affil-it]{authblk}
\usepackage{parskip}
\usepackage{textcomp}
\usepackage{csquotes}
\usepackage{color,soul}
\usepackage{xcolor}
\usepackage{multirow}
\usepackage{booktabs}

\usepackage[title]{appendix}
\usepackage{algorithm}
\usepackage{algorithmicx}
\usepackage{algpseudocode}
\usepackage{listings}
\usepackage{manyfoot}
\usepackage{tabularx}
\usepackage[dvipsnames]{xcolor}

\providecommand{\keywords}[1]
{
  \small	
  \textbf{\textit{Keywords---}} #1
}

\title{Trustworthy AI-based crack-tip segmentation using domain-guided explanations}

\author[1]{Jesco Talies}
\author[1]{Eric Breitbarth}
\author[1,*]{David Melching}

\affil[1]{Institute for Frontier Materials on Earth and in Space, German Aerospace Center (DLR), Linder Hoehe, Cologne, 51147, Germany}
\affil[*]{Corresponding author: David.Melching@dlr.de}

\date{\today}

\begin{document}
\maketitle

\begin{abstract}
Ensuring the trustworthiness and robustness of deep learning models remains a fundamental challenge, particularly in high-stakes scientific applications. In this study, we present a framework called \textit{attention-guided training} that combines explainable artificial intelligence techniques with quantitative evaluation and domain-specific priors to guide model attention. We demonstrate that domain-specific feedback on model explanations during training can enhance the model's generalization capabilities. We validate our approach on the task of semantic crack tip segmentation in digital image correlation data, which is a key application in the fracture mechanical characterization of materials. By aligning model attention with physically meaningful stress fields, such as those described by Williams’ analytical solution, attention-guided training ensures that the model focuses on physically relevant regions. This finally leads to improved generalization and more faithful explanations.
\end{abstract}


\section{Introduction}

\textbf{Deep learning} (DL) has led to enormous breakthroughs in many scientific fields, from computer vision~\cite{Voulodimos2018} to biology~\cite{Jumper2021}, material science~\cite{Schmidt2019, Merchant2023, Zeni2025}, computational mechanics~\cite{Herrmann2024},failure modeling~\cite{Aldakheel2021}, and fracture mechanics~\cite{S160} because of its ability to identify patterns in complex, high-dimensional data.
DL is based on the training of a highly flexible deep neural network, consisting of millions of trainable parameters, with large amounts of data. 
While typically improving performance, deep neural networks are trained end-to-end and lack interpretability and explainability.
This black-box problem raises critical questions regarding reliability and trustworthiness, in particular in high-risk and high-stakes applications such as autonomous vehicles and robots, healthcare, or maintenance of aircraft systems and components.
These considerations have been recently recognized as a legal issue in guidelines issued by the EU AI Act \cite{EuAiAct} and more specifically for aerospace applications by EASA \cite{EASA} and NASA \cite{NASA}.

\textbf{Explainable artificial intelligence} (XAI) \cite{BarredoArrieta2020} addresses these issues by proposing processes and methods that provide insights into the decision-making processes of DL models. There are two approaches to XAI: on the one hand, one can aim to achieve intrinsic interpretability by designing inherently transparent models that are human-understandable \cite{Zeng2016}. On the other hand, post-hoc explainability seeks to clarify model predictions without modifying the internal model structure, for example, through gradient-based sensitivity analysis \cite{Simonyan2013} or model-agnostic feature attribution methods \cite{SHAP, LIME}. While models designed for intrinsic interpretability offer clarity, they often lack the complexity needed to capture intricate data patterns, which can result in suboptimal performance. This trade-off between interpretability and accuracy is particularly evident in complex tasks tackled with DL, where simpler models do not suffice.

For \textbf{convolutional neural networks} (CNNs), a class of DL models used mainly for spatial, grid-like data, post-hoc explanations in the form of attention heatmaps based on class activation mappings (CAM) \cite{CAM} have gained wide popularity. While originally only designed for classification tasks, CAM-based methods have recently been extended to semantic segmentation architectures \cite{SegGradCAM} and applied for crack tip segmentation models \cite{MelchExplainCrack}. Despite their success, post-hoc explainability also faces criticism for producing explanations that can be misleading or unfaithful to the original black-box model or may oversimplify complex relationships \cite{StopExplain}. Moreover, even when focusing solely on CAM-based approaches, there is a vast array of methods,including gradient-based variants such as Grad-CAM and its extensions \cite{GradCAM, GradCAM++, GradCAM+, GradCAMEW}, as well as perturbation- and decomposition-based approaches \cite{ScoreCAM, EigenCAM, AblationCAM}, making it challenging to determine which method is best suited for a given model and task.

To address these concerns, researchers have called for rigorous, standardized metrics to quantitatively evaluate XAI methods. 
Vilone \& Longo~\cite{Vilone2021} propose a hierarchical classification of XAI methods and conceptualize their evaluation, also pointing out that research currently lacks a consensus on how to assess explainability.
Nauta et al.~\cite{Co12} propose conceptual properties such as correctness (i.e., faithfulness), completeness, and compactness to measure, among others, how well explanations align with the true behavior of a model and how effectively they convey relevant information to users.

Moreover, recent work has explored novel approaches that incorporate explanations into the training process. Some methods integrate human-based feedback on explanations, for example, through corrective supervision~\cite{Hint},
interactive rationale selection~\cite{Illume},
or explanation-informed learning objectives~\cite{XIL}.
These approaches demonstrate that interacting with and reflecting upon model explanations can prevent bias, improve accuracy, and reduce required training samples. However, relying on human interaction is tedious and time-consuming, and particularly for specific tasks, it requires experts with domain knowledge.
Recently, Stammer et al.~\cite{LSX} instead sourced explanatory feedback from a secondary critic model. Although their results show improved model generalization and provide more faithful explanations, it does not allow for the explicit inclusion of known concepts and domain knowledge.

\begin{figure}[h]
    \centering
    \includegraphics[width=\linewidth]{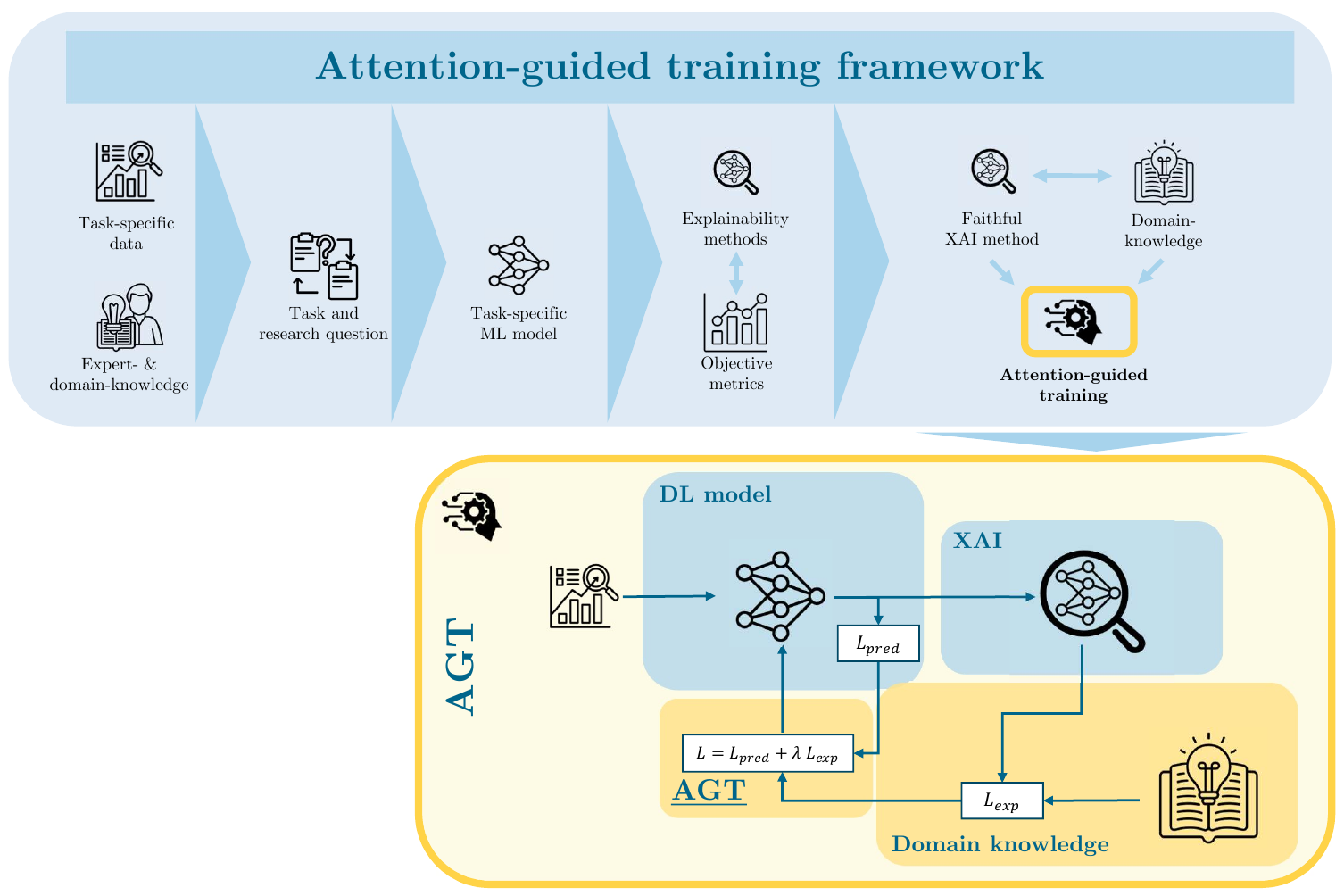}
    \caption{Overview of attention-guided training (AGT); AGT builds upon conventional machine learning principles, providing enhancements through explainable artificial intelligence (XAI) techniques. It incorporates post-hoc explanations and objective XAI evaluation metrics to assess model explainability and integrates domain knowledge through a modified training loop, ensuring better alignment between trained patterns and expert insights.
    }
    \label{fig:overview}
\end{figure}

Addressing these challenges, this paper introduces a framework that integrates faithful explanations and domain knowledge directly into the model training process. We call this framework \textbf{attention-guided training} (AGT). The framework is shown in Figure~\ref{fig:overview} and enables the determination of trustworthy explanations and provides alignment with domain knowledge during training. The framework starts with the identification of relevant expert knowledge for a specific task, which should be tackled by DL. Then, suitable explainability methods are identified and evaluated using objective metrics. Lastly, the core of AGT is a novel training process where the task-specific classical loss function is combined with an additional loss, which assesses the alignment of the explanation with domain-specific expectations.

\begin{figure}[h]
    \centering
    \includegraphics[width=\linewidth]{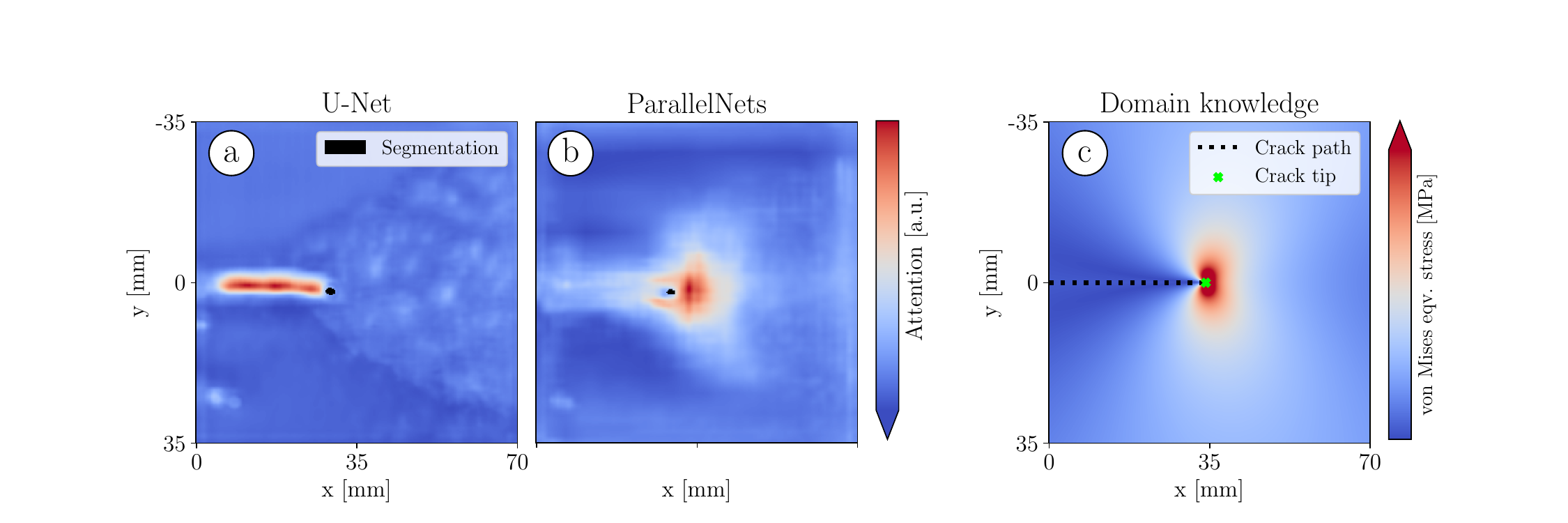}
    \caption{Semantic differences in model explanations, calculated using Grad-CAM, despite similar predictions.
    a) The explanation of the U-Net model prediction shows that the model identifies the crack path to predict the crack tip location. It is known that experimental data in this region likely contains artifacts from the digital image correlation evaluation (see Section \ref{sec:Data} for the experimental context).
    b) In contrast, the explanation of the ParallelNets \cite{MelchExplainCrack} model prediction highlights the region ahead of the crack tip as most relevant for its prediction 
    This attention aligns closely with domain knowledge shown in c), where the crack tip stress field is accurately represented by the widely accepted Williams series (Equation~\ref{eq:williams_stress}), thus relating model behavior directly to theoretical foundations in fracture mechanics.}
    \label{fig:motivation}
\end{figure}

While AGT serves as a general framework, its development was inspired by the findings of Melching et al. \cite{MelchExplainCrack}; hence, we focus on its specific application in the field of fracture mechanics. In \cite{MelchExplainCrack}, the authors employ the explainability method Grad-CAM \cite{GradCAM} for the semantic segmentation of fatigue crack tips in digital image correlation data. They train models with different architectures, including a U-Net \cite{UNet} and the so-called ParallelNets approach \cite{MelchExplainCrack}. The resulting model explanations exhibit distinct semantic characteristics, as illustrated in Figure \ref{fig:motivation}. While the U-Net primarily highlights the crack path, the ParallelNets explanations align more closely with the physical crack tip field described by Williams \cite{Williams1957}. Despite the improved generalization capabilities of the latter approach, a framework that explicitly controls network attention during training remains absent.

In this work, we present an attention-guided training of a U-Net model for crack tip segmentation in digital image correlation data. First, we adapt existing CAM-based XAI methods to semantic segmentation models. Next, we systematically evaluate their suitability for the given task by assessing correctness, completeness, continuity, and compactness, as proposed by Nauta et al. \cite{Co12}, and select the most appropriate XAI method accordingly. Finally, as AGT leverages domain knowledge to guide the training process and align model explanations with a predefined target, we calculate the representative von Mises equivalent stress field for the present load case — depicted exemplarily in Figure \ref{fig:motivation} c) — as the desired attention target. This choice is motivated by the findings of \cite{MelchExplainCrack} and discussions with experts in the field of fracture mechanics.

The paper is structured as follows:
In Section \ref{sec:results_xaieval}, we adapt CAM-based XAI methods to the application of crack tip segmentation in digital image correlation data and present their quantitative evaluation. 
Building on these evaluations, we apply AGT using physical target explanations motivated by domain knowledge in the form of the crack tip field. We then compare the resulting model performances with models trained using AGT on non-physical target explanations and a reference batch trained conventionally (without AGT) in Section~\ref{sec:results_agt}. 
In Section~\ref{sec:discussion}, we discuss the results together with possible applications and limitations of the framework.
Details regarding the methodology of the fracture mechanical experiments, the machine learning and XAI approach, and details on the AGT framework and application-specific implementations can be found in Section~\ref{sec:methods}.

\section{Results} \label{sec:results}
    We present the key findings of our study for the task of semantic segmentation of crack tips in DIC data. 
    
\begin{figure}
        \centering
        \includegraphics[width=\linewidth]{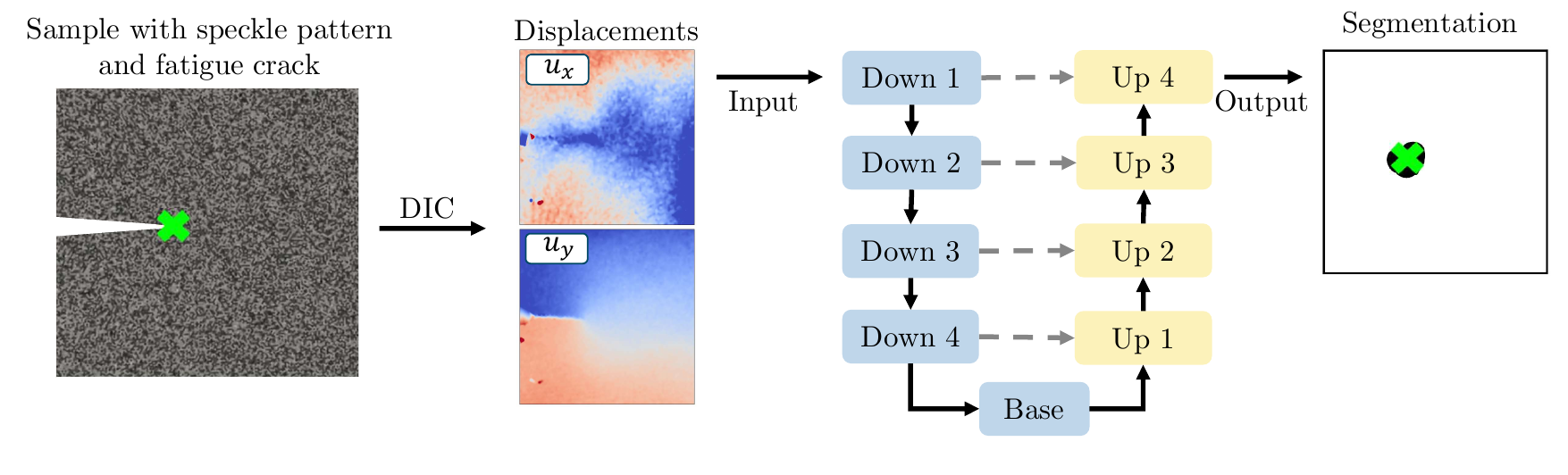}
        \caption{Illustration of the exemplary DL task. We obtain data through full-field DIC analysis during fatigue crack growth experiments (left). The resulting displacement fields are input into a segmentation CNN (here, U-Net architecture), which is trained to segment the most likely crack tip pixels. The centroid of the predicted segmentation yields the crack tip position, which is then translated back into the reference frame of the sample surface.}
        \label{fig:results_dltask}
    \end{figure}

    Accurately quantifying fatigue crack growth is essential for assessing the service life and damage tolerance of critical engineering structures and components exposed to variable service loads~\cite{tavares2017overview}.
    In recent years, digital image correlation (DIC) has played a crucial role in capturing full-field surface displacements and strains during fatigue crack propagation (FCP) experiments.
    The fracture mechanical evaluation of DIC data requires an exact and robust (i.e., reliable) determination of the crack tip positions~\cite{roux2009digital} -- an extremely challenging task due to inherent noise and artifacts~\cite{zhao2019state}. 
    Strohmann et al.~\cite{S160} created a labeled dataset and trained a U-Net~\cite{UNet} for crack tip segmentation. Melching et al.~\cite{MelchExplainCrack} refined this model and employed Grad-CAM~\cite{GradCAM} to generate attention heatmaps, providing explanations that guided the selection of models aligning with domain knowledge.
    An overview of the deep learning task is given in Figure~\ref{fig:results_dltask}.

    \subsection{Evaluation of explainability methods} \label{sec:results_xaieval}
\begin{figure}[ht]
            \centering
            \includegraphics[width=\linewidth]{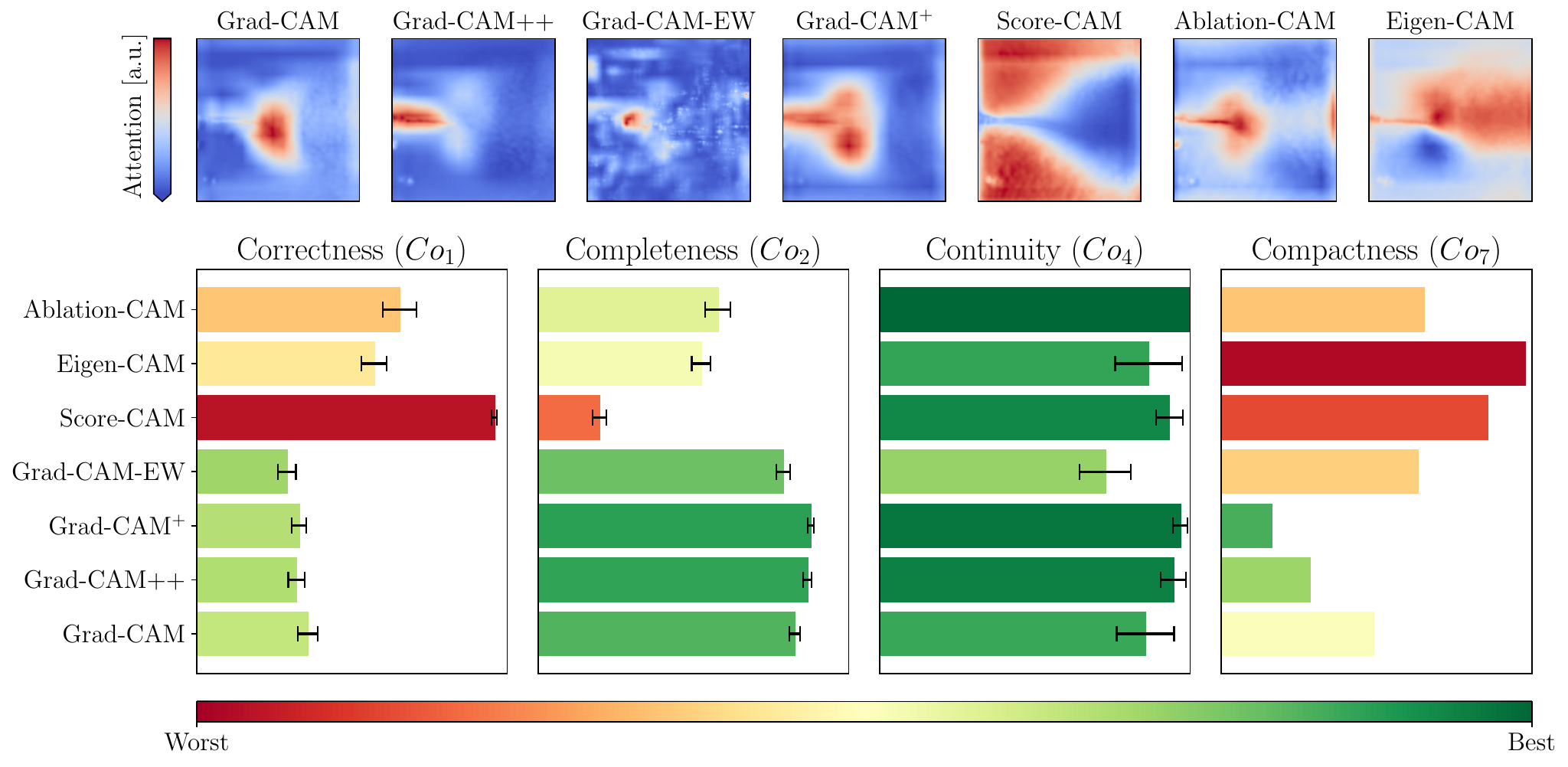}
            \caption{Visualization of explanations and performance comparison: The top row displays attention heatmaps generated using various CAM techniques, each corresponding to the same prediction made by \textit{ParallelNets} on a sample of the \(S_{160,4.7,val}\) dataset (see Section \ref{sec:Data}). These heatmaps are based on the encoder layers of the U-Net architecture. Given the significant visual differences between methods, the accompanying bar plot quantifies their relative performance based on the four criteria --- correctness, completeness, continuity, and compactness.}
            \label{fig:co12eval}
        \end{figure}

        In order to answer the question of which explainability method to choose, we fix the model choice to a single trained instance, namely the open-source \textit{ParallelNets} from Melching et al.~\cite{MelchExplainCrack}, truncated to the U-Net architecture for inference. 
        We generate explanations for various methods based on class activation mapping (CAM). We refer to Appendix~\ref{app:XAI} below for details on these methods and their adaptation to semantic segmentation.
        The resulting explanations in the form of attention heatmaps for the different methods are shown in Figure~\ref{fig:co12eval} (top) and show significant differences, both regarding semantic concepts and intuitive quality.
        While minor differences are expected and can be observed in explanations generated for classification CNNs as well, see, e.g., \cite{ScoreCAM, LayerCAM}, the adaptation to semantic segmentation appears to have a significant impact. Specifically, the explanations differ considerably across methods, in contrast to the more consistent patterns observed for classification in \cite{CompareCAM}.
        We further observed that the choice of considered neural block(s) and score (see Figure~\ref{fig:overview} and Appendix~\ref{app:XAI}, respectively) impact the resulting attention heatmaps even within the context of a single CAM method.
        This is related to the different fidelity and nature of features learned by each model block.
        Furthermore, we hypothesize that the skip connections in the U-Net architecture, along with the associated feature propagation, further contribute to a more complex and less localized feature representation \cite{veit2016residual}. 
        As a result, depending on the explanation method, this complexity may not be adequately captured in the final explanations.
        Since each of the considered methods aims to explain the predictions of the underlying model, our objective is to quantitatively assess which method is best suited for this task. 
        To achieve this, we employ a subset of the Co-12 criteria proposed by Nauta et al.~\cite{Co12}.
        We objectively evaluate which of the provided methods is the most faithful, referred to, in the spirit of Co-12, as correctness (\(Co_1\)), complete (\(Co_2\)), continuous (\(Co_4\)), in the sense that small changes in the input data lead to small changes in the explanations, and compact (\(Co_7\)), referring to the size of the explanation.
        Using data obfuscation strategies, we implemented metrics for each of these four criteria and refer to Section~\ref{sec:Co12} for details.
        Figure~\ref{fig:co12eval} (bottom) presents a concise, comprehensive comparison of the method performance among these criteria.
        
        For the chosen \textit{ParallelNets} model and the task of crack tip segmentation, the above metrics conclusively indicate that the gradient-based methods (Grad-CAM~\cite{GradCAM}, Grad-CAM++~\cite{GradCAM++}, Grad-CAM\(^+\)~\cite{GradCAM+}, and elementwise-GradCAM~\cite{GradCAMEW}) are significantly more \textit{correct, complete,} and \textit{compact}, but are similarly \textit{continuous}, compared to the gradient-free approaches (Score-CAM~\cite{ScoreCAM}, Eigen-CAM~\cite{EigenCAM}, and Ablation-CAM~\cite{AblationCAM}).
        While, e.g., Score-CAM appears unsuitable for this model and task, particularly due to its poor performance in terms of correctness and completeness, Grad-CAM+ and Grad-CAM++ demonstrate strong overall evaluation results.
        Ultimately the results found in Figure~\ref{fig:co12eval} indicate that it is entirely possible to explain segmentation models using the CAM techniques originally intended for classification CNNs.

    \subsection{Attention-guided training (AGT)} \label{sec:results_agt}
        Considering the vast possibilities of model attention patterns, as presented in Figure~\ref{fig:motivation}, we argue that it is beneficial to guide these patterns by aligning them with domain-specific, theory-guided target explanations.
        The alignment is carried out by training a model with a total loss function \( L_{\rm total} \), combining a prediction loss \(L_{\rm pred}\) and an explanation loss \( L_{\rm exp} \)  between the current and target explanation.
        We provide the target explanation $\mathbf{\hat \Phi}$ using the relevant domain knowledge. The total loss is expressed as:
        \begin{equation}
            \label{eq:AGTloss}
            L_{\rm total} = L_{\rm pred}(\mathbf{y},\mathbf{\hat{y}}) + \lambda L_{\rm exp}(\mathbf{\Phi},\mathbf{\hat\Phi}),
        \end{equation}
        where \(\lambda \ge 0\) is a hyperparameter balancing both loss contributions. Here, \(\mathbf{y}\) and \(\mathbf{\hat{y}}\) denote the actual and target predictions, respectively, while \(\mathbf{\Phi}\) and \(\mathbf{\hat \Phi}\) denote the current and domain-guided target explanations, respectively.
        Thus, the explanation loss term ensures coherence (\(Co_{11}\)) of explanations with expert knowledge (see Section~\ref{sec:Co12}).
        
        For the case of crack tip segmentation, this total loss consists of the Dice loss~\cite{sudre2017generalised} between current and target predictions and the cosine similarity (\(S_C\)) between current and target explanations.
        The loss then reads:
        \begin{equation}
            \label{eq:AGTloss_crack}
            L_{\rm total} = \textrm{Dice}(\mathbf{y},\mathbf{\hat{y}}) + \lambda S_{C}(\mathbf{\Phi},\mathbf{\hat \Phi}),
        \end{equation}
        where
        \begin{equation*}
            \textrm{Dice}(\mathbf{y},\mathbf{\hat{y}}) = 1 - \frac{2 \sum_{ij} y_{ij} {\hat y}_{ij}}{\sum_{ij}(y_{ij} + {\hat y}_{ij})},
        \end{equation*}
        and
        \begin{equation*}
            S_{C}(\mathbf{\Phi},\mathbf{\hat \Phi}) = 1 - \frac{\sum_{ij} \phi_{ij} {\hat \phi}_{ij}}{\big(\sum_{ij}\phi_{ij}^2\big)^{1/2} \big(\sum_{ij}{\hat \phi}_{ij}^2\big)^{1/2}}.
        \end{equation*}
        
        We further evaluate the model's generalization capabilities in terms of reliability. Reliability scores serve as a quantitative measure of generalization performance and are defined as the fraction of samples with valid segmentations (i.e., exactly one contiguous patch of segmented pixels; Section~\ref{sec:Loss}) for any given dataset.
    \subsubsection*{Training phases} \label{sec:results_trainingphases}
        To ensure successful training, we divide the process into two phases. The first phase, an initial pretraining, follows a conventional DL approach using only the prediction loss (i.e., \(\lambda = 0\)), which serves to prime the explanatory component of AGT and allows the model to produce meaningful explanations. In the second phase, a finite \(\lambda > 0\) is introduced to refine the learned behavior towards the theory-guided target explanations \(\mathbf{\hat \Phi}\). To preserve the model's predictive performance throughout the alignment stage, the hyperparameter \(\lambda\) must remain sufficiently small, preventing over-steering, which would result in non-salient explanations. For this particular use case, the useful range $\lambda\in (0.5,3)$ has been empirically estimated.
    
    \subsubsection*{Domain knowledge} \label{sec:results_domain}
        For crack tip segmentation, our domain knowledge is based on analytical expressions describing the displacement and stress fields near the crack tip of an open crack.
        More precisely, we use the analytical derivation by Williams~\cite{Williams1957}:
        
        In planar linear-elastic fracture mechanics, the stress and displacement fields induced by a single open crack with traction-free crack faces can be described in polar coordinates \((r, \theta)\) by the Williams series expansion \cite{Kuna}
        \begin{align}
            \sigma_{ij}(r,\theta) &= \sum_{n} {r^{\frac{n}{2}-1}\ \left( A_n f_{\text{I},ij}(\theta,n) + B_n f_{\text{II},ij}(\theta,n) \right)}, \label{eq:williams_stress} \\
            u_{ij}(r,\theta) &= \sum_{n} {\frac{r^{\frac{n}{2}}}{2\mu}\ \left( A_n g_{\text{I},ij}(\theta,n) + B_n g_{\text{II},ij}(\theta,n) \right)}. \label{eq:williams_displ}
        \end{align}
        To estimate the stress field \(\sigma_{ij}\) in Equation \eqref{eq:williams_stress} for experimental DIC data, we optimize the parameters \(A_n, B_n \in \mathbb{R}\), called Williams coefficients, by fitting the theoretical field \eqref{eq:williams_displ} to the DIC displacement data using the over-deterministic method implemented in CrackPy~\cite{Crackpy}.
        From this, the von Mises equivalent stress is computed as $\sigma_{\rm VM} = \sqrt{\sigma_{11}^2 + \sigma_{22}^2 - \sigma_{11}\sigma_{22} + 3\sigma_{12}^2}$.
    
    \subsubsection*{Effectiveness of AGT} \label{sec:results_exampleagttrain}
\begin{figure}
            \centering
            \includegraphics[width=1.0\linewidth]{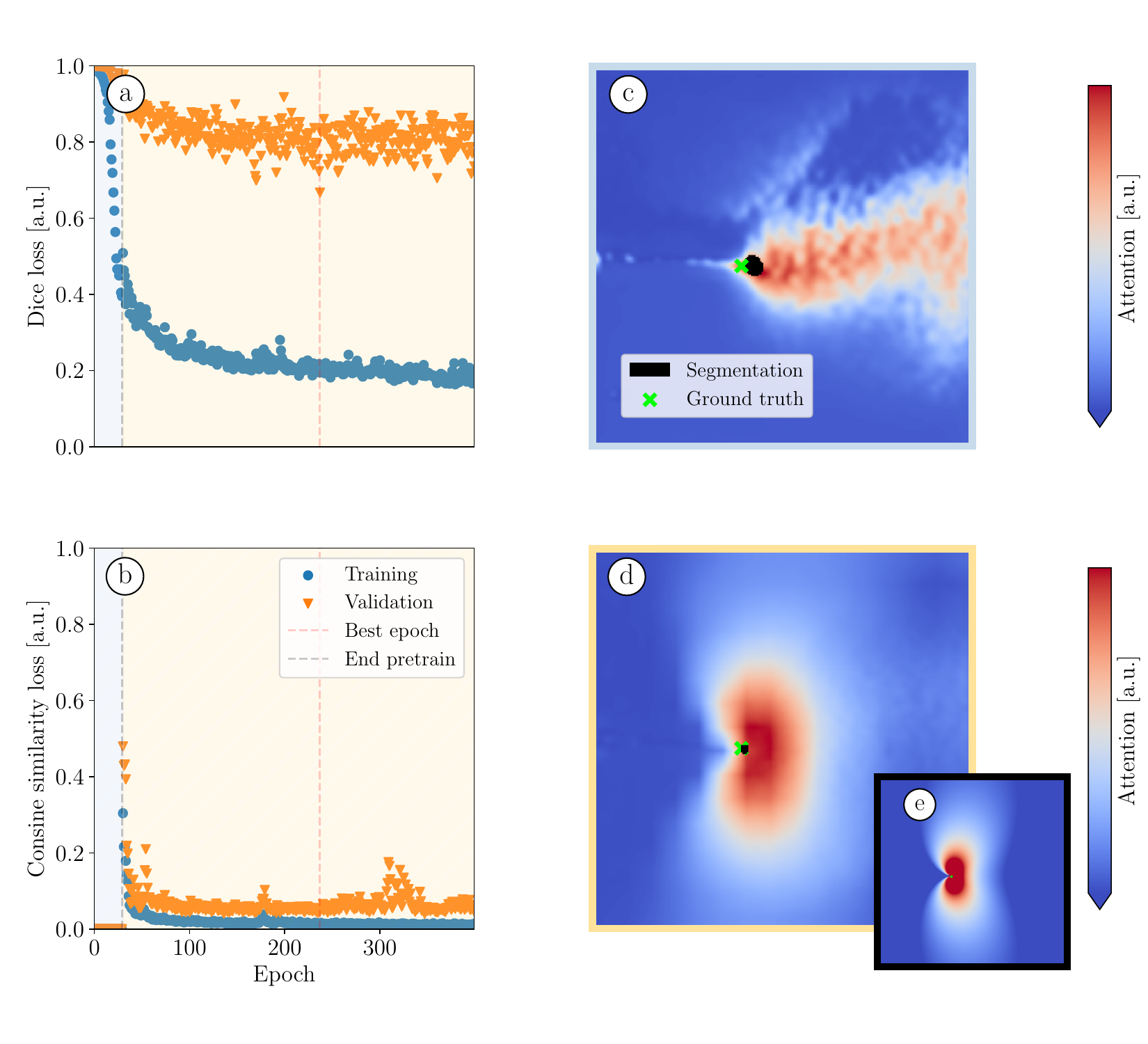}
            \caption{
            Example of an AGT for crack tip segmentation. The Dice loss provides feedback on the segmentation quality, while the cosine similarity loss quantifies coherence with the target explanation (e). 
            Plots (a) and (b) show the loss curves of the segmentation quality and explanation coherence, respectively.
            Plot (c) shows the attention heatmap after the pretraining phase of 30 epochs.
            Plot (d) depicts the heatmap of the model with the smallest validation loss during the AGT phase (indicated by a dashed orange line in (a, b)).
            All attention heatmaps were obtained using the Grad-CAM++ method on the encoder layers of the U-Net.}
            \label{fig:TrainingLoss}
        \end{figure}

        Figure~\ref{fig:TrainingLoss} shows an example result of AGT for crack tip segmentation, illustrating the initial unaligned attention after 30 epochs of pretraining (see Figure~\ref{fig:TrainingLoss} a), c)) and the final explanation, aligned to the theory-guided target (see Figure~\ref{fig:TrainingLoss} e)). The number of epochs used for pretraining was empirically determined. To avoid overfitting, we used the trained weights saved from the epoch with the lowest total validation loss (see Figure~\ref{fig:TrainingLoss} b), d)) as the final model to compute predictions and corresponding explanations.
        In this example, the target explanation is based on the corresponding stress field using the strategy Gradual Williams (GW) introduced below. This strategy steers the attention towards regions of high von Mises equivalent stress larger than 75 MPa (see Figure~\ref{fig:TrainingLoss} e)). For details on the used processing of our target explanations, we refer to Section~\ref{sec:AGT} and Appendix~\ref{app:AGT}.
        
        Preliminary experiments indicated that it is beneficial to avoid large corrective updates; large $\lambda$ can cause over-correction of the model weights within the first few AGT epochs, which causes the predictive performance to deteriorate, leading to non-salient explanations. Models found in this state were rarely able to recover the intended training. Similarly, small $\lambda$ had insufficient effect on the resulting explanations. To mitigate these effects, loss scaling was applied, and the present experiment was conducted using a weight factor of $\lambda = 2$.
        We observe that both training phases, pretraining and attention-guided, successfully converge. In this example, both validation and explanation loss exhibit significant variance, underpinning our approach of selecting the epoch with the lowest validation loss. For this model, the attention (Figure~\ref{fig:TrainingLoss} d)) aligns with the target Gradual Williams attention (Figure~\ref{fig:TrainingLoss} e)), demonstrating the effectiveness of AGT.

    \subsubsection*{Comparison of attention strategies} \label{sec:attentionStrats}
        To  investigate whether models guided by physically meaningful explanations exhibit improved generalization and trustworthiness, we conduct a series of experiments. For this, we introduce different target attention maps -- two \textit{physical} strategies that build on expert knowledge and intuition, i.e.,
    
        \begin{itemize}
            \item \textit{Binary Williams (BW)}: The target explanation is obtained by binarizing the Williams stress field using a fixed threshold: regions of elevated mechanical stress were assigned a target attention of 1, all others 0,
            \item \textit{Gradual Williams (GW)}: The target explanation is derived by truncating the Williams stress field at a specified threshold and rescaling the resulting values to the range \([0, 1]\), yielding a continuous, non-binary attention map with gradually fading intensity,
        \end{itemize}
        
        and two \textit{non-physical} strategies that are intentionally designed to steer attention towards allegedly less informative regions, i.e.,
        
        \begin{itemize}
            \item \textit{Binary misleading (BM)}: The target attention is set to 1 within a small square located in the bottom-right corner of the domain and 0 elsewhere.
            \item \textit{Multi-gradual misleading (MGM)}: The target attention is set to 1 at the top- and bottom-right corners and gradually fades to 0.
        \end{itemize}
    
\begin{figure}
            \centering
            \includegraphics[width=\linewidth]{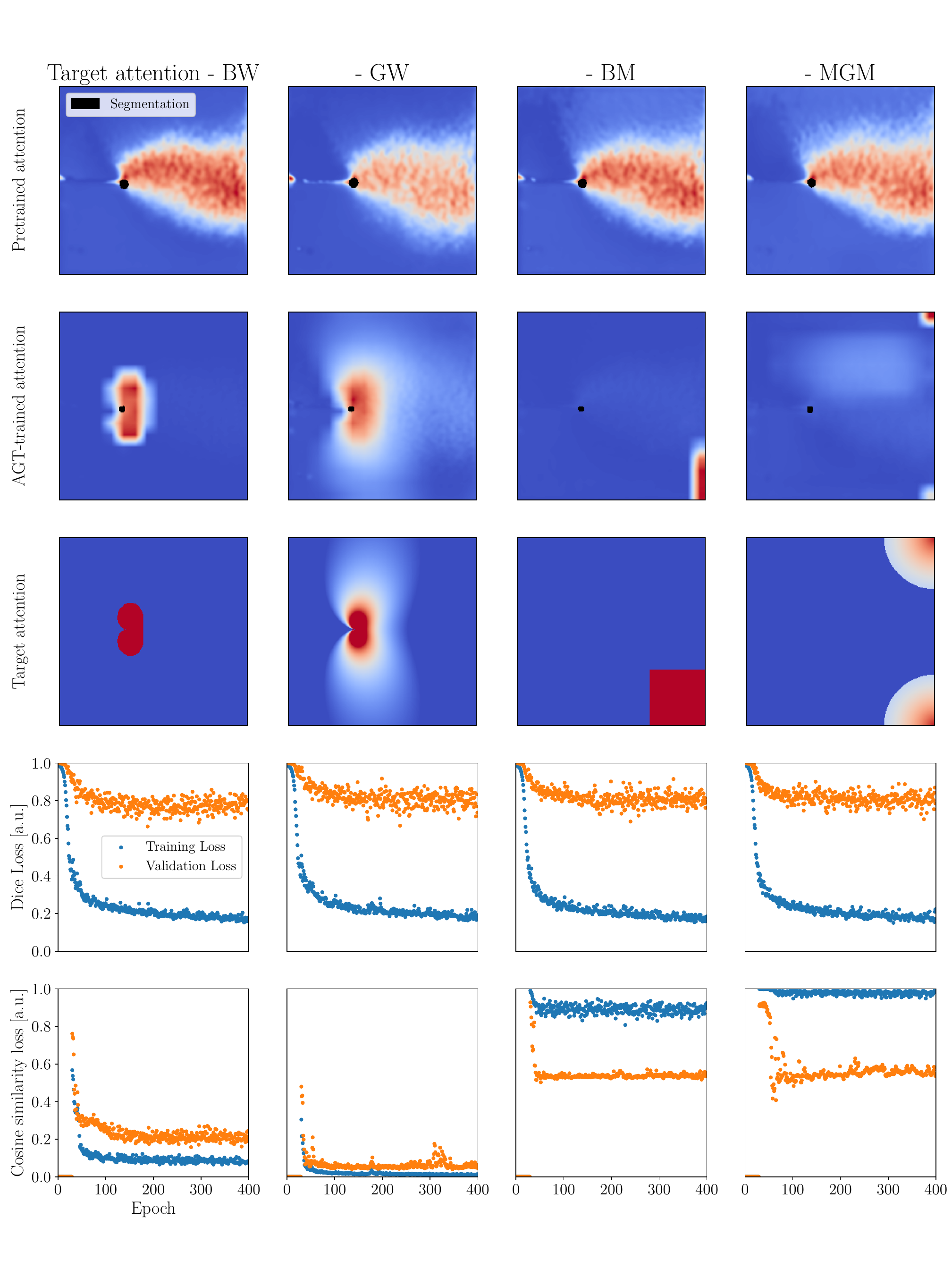}
            \caption{
            Comparison of different attention target strategies for exemplary training runs: \textit{Binary Williams} (BW) and \textit{Gradual Williams} (GW) are adaptations of the Williams stress field, whereas \textit{Binary Misleading} (BM) and \textit{Multi-Gradual Misleading} (MGM) are intentionally steering attention towards allegedly less informative regions.
            All attention heatmaps were obtained using the Grad-CAM++ method on the encoder layers of the U-Net.
            }
            \label{fig:TrainingVariations}
        \end{figure}

        The results presented in Figure \ref{fig:TrainingVariations} show similar features as discussed in more detail in Figure \ref{fig:TrainingLoss}.
        Although the models can be qualitatively aligned with each of the considered explanation strategies without compromising predictive performance, the quantitative alignment of the explanation loss term is markedly lower for the non-physical target attentions compared to those guided by more salient, physically informed strategies.

        Considering the large volatility in the validation loss and following up on previous work \cite{MelchExplainCrack}, we evaluate the models trained with different AGT strategies regarding their robustness and trustworthiness. For this, we compare the validation (Dice) loss, the reliability of the model (see Section~\ref{sec:Loss}), and the explanation correctness (see Section~\ref{sec:Co12}), ensuring that the learned explanations are faithful.
        By calculating both validation loss and reliability on in-distribution and reliability on out-of-distribution datasets (see Section~\ref{sec:Data}), we can qualitatively estimate the generalization capabilities of the respective models.

        For each attention strategy, we trained 10 randomly initialized models. All experiments were performed with \(\lambda = 2\), which was determined empirically and worked consistently for the present task. In addition, we trained 10 models without AGT as an independent baseline (Reference (R); $\lambda = 0$).
        
\begin{figure}
            \centering
            \includegraphics[width=\linewidth]{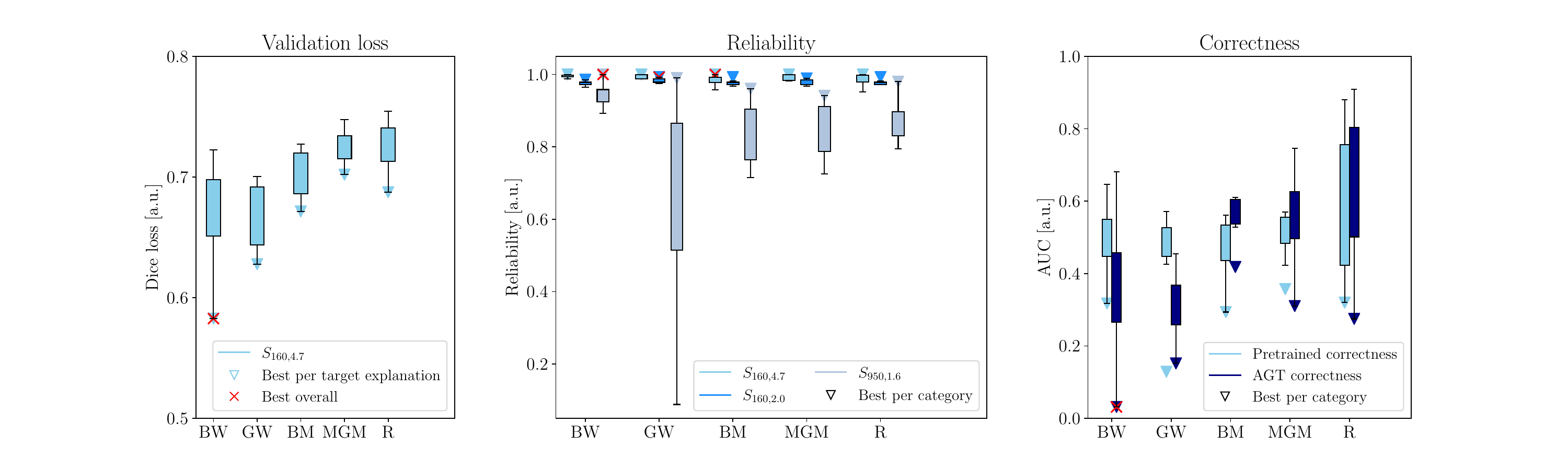}
            \caption{
            Quantitative evaluation of attention strategy experiments.
            Left: Validation (Dice) loss (lower is better) to measure predictive performance. Middle: Reliability of detection (larger is better) as a metric for generalization capacity (see Section~\ref{sec:Loss}). Right: Explanation correctness, measured by area under the curve (AUC) (lower is better) to characterize faithfulness of explanations (see Section~\ref{sec:Co12}). Both in-distribution (\(S_{160,4.7}\)) and out-of-distribution datasets (\(S_{160,2.0}, S_{950,1.6}\)) are considered (see Section~\ref{sec:Data}). Numeric representation can be found in Table~\ref{tab:Results}.
            }
            \label{fig:Results}
        \end{figure}

        The results of this study are presented visually in Figure~\ref{fig:Results} and quantitatively in Table~\ref{tab:Results}, with statistical significance assessed using Mann–Whitney-U tests (Appendix~\ref{app:stats}).
        
        On average, models trained with AGT using physical target explanations (BW, GW) achieve lower validation loss compared to the non-physical (BM, MGM) and the unguided reference (R). This difference is statistically significant when comparing physical strategies against both misleading targets  and the reference baseline (see Table~\ref{tab:mwu_results}; Appendix~\ref{app:stats}).
        Among all strategies, the binary Williams (BW) explanation yields the best single model in terms of validation loss. 
        
        In terms of reliability, on the in- or near-distribution datasets (\(S_{160,4.7}, S_{160,2.0}\)), all strategies exhibit saturated behavior, whereas for the further out-of-distribution datasets (\(S_{950,1.6}\)) the binary Williams (BW) strategy shows significantly higher reliability values compared to every other strategy with statistical significance (see Appendix~\ref{app:stats} for details).

        As far as correctness is concerned, the physical strategies (BW and GW) improve the correctness of their explanation with AGT, while non-physical strategies (BM and MGM) deteriorate correctness, as indicated by decreased and increased Area Under the Curve (AUC) values, respectively. Moreover, the physical strategy (BW) resulted in the overall best model in terms of validation loss, reliability, and correctness of the explanations.

\begin{table}[t]
\centering
\caption{Performance comparison across target explanation types. Reported values are mean $\pm$ standard deviation and 95\% confidence intervals over 10 independent training runs. The evaluate model instances were selected based on the lowest-achieved validation loss, representing their respective peak performance potential. Sub-tables show (a) validation loss on the in-distribution dataset $S_{160,4.7}$; (b) reliability on in-distribution and out-of-distribution datasets; and (c) explanation correctness before and after attention-guided training (AGT). Bold values indicate the best-performing method within each category. For correctness, additionally the relative change induced by AGT is evaluated per method. Colored arrows indicate whether explanation trustworthiness improved ($\downarrow$), deteriorated ($\uparrow$), or remained unchanged (--).}
\begin{subtable}[t]{\textwidth}
\centering
\caption{Validation loss (lower is better)}
\begin{tabular}{l || c  c}
\hline
 & \multicolumn{2}{c}{$S_{160,4.7}$} \\
 
 & \multicolumn{1}{c}{Mean $\pm$ Std}
 & \multicolumn{1}{c}{95\% CI} \\
\hline
BW  & \textbf{0.67 $\pm$ 0.04} & [0.64, 0.69] \\
GW  & \textbf{0.67 $\pm$ 0.03} & [0.65, 0.68] \\
BM  & 0.70 $\pm$ 0.02 & [0.69, 0.71] \\
MGM & 0.72 $\pm$ 0.01 & [0.72, 0.73] \\
R   & 0.72 $\pm$ 0.02 & [0.71, 0.74] \\
\hline
\end{tabular}
\end{subtable}

\vspace{0.8em}

\begin{subtable}[t]{\textwidth}
\centering
\caption{Reliability (higher is better)}
\begin{tabular}{l || c c | c c | c c}
\hline
 & \multicolumn{2}{c}{$S_{160,4.7}$} & \multicolumn{2}{c}{$S_{160,2.0}$} & \multicolumn{2}{c}{$S_{950,1.6}$} \\
 
 & \multicolumn{1}{c}{Mean $\pm$ Std}
 & \multicolumn{1}{c}{95\% CI}
 & \multicolumn{1}{c}{Mean $\pm$ Std}
 & \multicolumn{1}{c}{95\% CI}
 & \multicolumn{1}{c}{Mean $\pm$ Std}
 & \multicolumn{1}{c}{95\% CI} \\
\hline
BW & \textbf{0.99 $\pm$ 0.01} & [0.99, 1.00] & 0.97 $\pm$ 0.01 & [0.97, 0.98] & \textbf{0.94 $\pm$ 0.05} & [0.90, 0.96] \\
GW & \textbf{0.99 $\pm$ 0.02} & [0.97, 1.00] & \textbf{0.98 $\pm$ 0.01} & [0.98, 0.99] & 0.67 $\pm$ 0.28 & [0.50, 0.82] \\
BM & 0.98 $\pm$ 0.01 & [0.98, 0.99] &  \textbf{0.98 $\pm$ 0.01} & [0.97, 0.98] & 0.78 $\pm$ 0.22 & [0.63, 0.88] \\
MGM & 0.98 $\pm$ 0.03 & [0.97, 1.00] & \textbf{0.98 $\pm$ 0.01} & [0.97, 0.98] & 0.85 $\pm$ 0.08 & [0.80, 0.90] \\
R & \textbf{0.99 $\pm$ 0.02} & [0.98, 0.99] & \textbf{0.98 $\pm$ 0.01} & [0.97, 0.98] & 0.86 $\pm$ 0.12 & [0.79, 0.92] \\
\hline
\end{tabular}
\end{subtable}

\vspace{0.8em}

\begin{subtable}[t]{\textwidth}
\centering
\caption{Correctness (lower is better)}
\begin{tabular}{l || c c | c c | c}
\hline
 & \multicolumn{2}{c}{Pretrained - $S_{160,4.7}$}
 & \multicolumn{2}{c}{AGT-trained - $S_{160,4.7}$} \\
 
 & \multicolumn{1}{c}{Mean $\pm$ Std}
 & \multicolumn{1}{c}{95\% CI}  
 & \multicolumn{1}{c}{Mean $\pm$ Std}
 & \multicolumn{1}{c}{95\% CI} \\

\hline
BW  & 0.51 $\pm$ 0.1 & [0.45, 0.57] & 0.36 $\pm$ 0.18 & [0.25, 0.46] & $\textcolor{OliveGreen}{\downarrow}$ \\
GW  & 0.47 $\pm$ 0.14 & [0.38, 0.54] & 0.31 $\pm$ 0.09 & [0.25, 0.37] & $\textcolor{OliveGreen}{\downarrow}$ \\
BM  & 0.47 $\pm$ 0.10 & [0.41, 0.52] & 0.59 $\pm$ 0.12 & [0.52, 0.66] & $\textcolor{Maroon}{\uparrow}$ \\
MGM & 0.52 $\pm$ 0.09 & [0.46, 0.57] & 0.55 $\pm$ 0.13 & [0.47, 0.62] & $\textcolor{Maroon}{\uparrow}$ \\
R   & 0.60 $\pm$ 0.22 & [0.49, 0.72] & 0.60 $\pm$ 0.22 & [0.49, 0.72] & $\textcolor{Black}{-}$\\
\hline
\end{tabular}
\end{subtable}

\label{tab:Results}
\end{table}

\section{Discussion} \label{sec:discussion}
    Utilizing deep learning (DL) methods in a faithful and trustworthy manner remains a challenge, especially in scientific domains, as highlighted by recent work on robustness and generalization~\cite{Sapoval2022} and on the faithfulness of learned representations~\cite{Scorzato2024}.
    In this work, we show that integrating explainable artificial intelligence (XAI) techniques with evaluation metrics and domain knowledge to guide model attention can enhance both generalization capabilities and trustworthiness of DL models. We validate this claim in the context of machine-learned crack tip segmentation in full-field displacement fields obtained by digital image correlation during fatigue crack growth, which is a critical task in fracture mechanics where model interpretability and robustness are paramount.
    
    We build upon state-of-the-art methodology by extending techniques based on class activation mapping (CAM) to semantic segmentation tasks, similar to \cite{SegGradCAM}.
    These methods can be applied to a variety of neural layers of the network, or a combination of such, and provide meaningful insight into the internal decision-making process of our DL model, as illustrated in Figures~\ref{fig:motivation} and \ref{fig:co12eval}.
    Adapting these methods has proven to be non-trivial. Even upon fixing the CAM target layers, the visualizations in Figure~\ref{fig:co12eval} reveal substantial variation among different CAM-based methods with respect to attribution shape, spatial alignment, and total relevance.
    These differences underscore the heuristic and often inconsistent nature of post-hoc XAI techniques, exposing them to warranted critique \cite{StopExplain} and necessitating cautious application, especially in contexts where explanation fidelity (correctness \cite{Co12}) is critical.
    
    Therefore, it is imperative to complement the visualizations with objective and quantitative metrics, evaluating, among others, the fidelity between the model and provided explanations.
    To that end, we address four of the twelve XAI evaluation criteria of Nauta et al.~\cite{Co12}, tailored to the task of crack tip segmentation in digital image correlation data.
    Evaluation of these metrics allows us to systematically determine the faithfulness and quality of different methods, as depicted in Figure~\ref{fig:co12eval}, identifying the gradient-based methods Grad-CAM \cite{GradCAM} and Grad-CAM++ \cite{GradCAM++} as most effective for this task and data. 
    While these objective criteria provide valuable insights, further investigation is required to refine the implementation of the metrics and their applicability to other tasks and explanation methods. In general, it should be mentioned that the evaluation results depend on multiple parameters, among others the chosen model architecture, task, and data. Therefore, one cannot provide general guidance or static performance values, as each use case and model has to be evaluated separately.
    
    To leverage explanations during training, we adapt the Learning by Self-Explaining (LSX) approach recently proposed by Stammer et al. \cite{LSX}. LSX introduces a novel training paradigm where a learner model is optimized not only for the primary predictive task but also through feedback from a critic model that evaluates the quality of the determined explanations.
    However, a limitation of the LSX framework is that it relies on explanations that have not been externally validated.
    
    In our attention-guided training (AGT) approach, we address these issues by choosing a correct, complete, continuous, and compact explanation method (see Figure \ref{fig:co12eval}) and incorporating domain-specific knowledge to inform and evaluate explanations, thereby providing more reliable feedback and enhancing the overall trust in the model's predictions and explanations.
    This idea is motivated by earlier work conducted by Melching et al.~\cite{MelchExplainCrack}, where the authors observed that different model architectures lead to distinct attention patterns. Specifically, while some models learn to focus attention on physically relevant regions like the crack path or crack tip field, others display attention in presumably less meaningful areas, indicating that not all trained models inherently learn the same physical features. However, the combination of physical features and the inherent flexibility of deep learning often makes these models more powerful in terms of generalization capabilities.
    To guide model attention towards regions of physical significance, i.e., areas of high mechanical stress, we utilize the well-known von Mises stress field calculated using a Williams series expansion fitted to displacement data as the physical attention prior.
    
    Our findings show that models guided away from the natural crack path attention -- including the non-physical strategies (BM and MGM) -- exhibit predictive performance comparable to, or slightly exceeding, that of the unguided reference model.
    In this study, predictive performance is assessed using a qualitative combination of the validation Dice coefficient, computed on the spatially and statistically distinct validation side of our labeled dataset $S_{160,4.7}$, and a task-specific reliability metric that enables performance estimation on unlabeled, out-of-distribution datasets $S_{160,2.0},\ S_{950,1.6}$. 
    
    Due to the severe class imbalance of the segmentation task and the resulting stochastic fluctuations during training, model instances are selected based on their peak validation Dice performance. Consequently, the quantitative results reported in Figure~\ref{fig:Results} and Table~\ref{tab:Results} represent the maximum achievable performance of each training strategy rather than an unbiased estimate of expected deployment behavior. While this model selection procedure constitutes a known form of optimization bias, it is applied uniformly across all training regimes. As a result, relative comparisons between strategies remain meaningful and allow a fair assessment of their achievable performance.
    
    Within this evaluation protocol, physically inspired attention strategies (BW and GW) consistently outperform both misleading and unguided baselines. This is reflected in lower mean validation Dice losses ($\approx 0.67$) compared to misleading and reference strategies ($\geq 0.7$; Table~\ref{tab:Results}).
    One-sided non-parametric comparisons using the Mann–Whitney U test confirm that physically guided strategies yield significantly lower Dice losses than misleading baselines ($p < 10^{-5}$) and the unguided reference ($p < 10^{-3}$); full test statistics are reported in Appendix~\ref{app:stats}.
    
    We attribute the improved performance of physically guided models to the induction of information-dense, crack-tip-field–dependent decision patterns in the model’s internal representation. The marginal performance gains of the misled models over the unguided reference are likely attributed to the avoidance of experimental artifacts inherent to digital image correlation measurements near the crack path. All guided strategies -- physical and non-physical alike -- systematically suppress attention in this noise-dominated region and instead emphasize areas containing crack-tip field information (see Figure~\ref{fig:TrainingVariations}), which appears sufficient to recover modest segmentation gains even in the absence of physically meaningful guidance.
    
    Beyond in-distribution performance, generalization to unlabeled out-of-distribution data can be assessed using the task-specific reliability metric. Here, physically guided models exhibit markedly improved robustness, most prominently on the non-saturated OOD dataset $S_{950,1.6}$.
    In this regime, the Binary Williams (BW) strategy achieves mean reliability scores of approximately $94\%$, whereas all other strategies do not exceed $86\%$. 
    Corresponding hypothesis tests confirm that BW yields significantly higher reliability scores than each other attention strategy ($p < 7\cdot 10^{-3}$; Appendix~\ref{app:stats}). Evidently, providing a binarized, physically grounded representation of the crack-tip field as an attention prior is particularly effective in promoting reliable and robust model behavior under distributional shift.
    
    Importantly, even under this favorable peak-performance evaluation protocol, which affords all models their best possible chance, physically guided strategies retain a clear and statistically supported advantage. This demonstrates that aligning model attention with domain-consistent stress-field priors raises the attainable performance beyond what can be achieved through unguided or arbitrarily guided training.
    
    With respect to explanation correctness, physically guided strategies (BW and GW) exhibit a consistent improvement after attention-guided training, whereas non-physical strategies (BM and MGM) show a deterioration in correctness, as indicated by decreasing and increasing area under the curve (AUC) values, respectively (see Table~\ref{tab:Results}c)). Importantly, this trend cannot be attributed to longer training alone: although all AGT-trained models undergo an extended optimization phase, the correctness of the unguided reference model ($\lambda=0$) remains unchanged throughout the training, while correctness for misleading strategies degrades. This demonstrates that the observed correctness changes are not an artifact of additional training epochs but are specifically induced by attention guidance.
    
    The pretrained correctness values of all strategies exhibit noticeable variability; however, these initial values lie comfortably within the standard deviation of the reference baseline ($0.60 \pm 0.22$) and can therefore be attributed to the inherent stochasticity of deep learning training rather than systematic differences between strategies. The subsequent divergence in correctness-improvement for physically guided models and deterioration for misleading ones thus reflects a genuine effect of the respective attention strategies.
    
    While, in principle, predictive accuracy and explanation quality need not be coupled, our results indicate that under attention-guided training, aligning optimization with physically meaningful explanations can positively influence explanation faithfulness. In particular, models guided by physical attention priors consistently produce more faithful explanations, suggesting a non-trivial interplay between prediction learning and explanation alignment. This interaction represents a promising direction for future investigation into training paradigms that jointly optimize predictive performance and interpretability.
    
    This work presents a framework for objectively selecting suitable XAI methods for specific semantic segmentation tasks and introduces a novel approach to guide model attention by domain knowledge.
    It serves as a compelling example of how XAI can move beyond post-hoc interpretation to actively inform and guide model development. 
    Importantly, the proposed framework is not limited to specific data types, physical models, or explanation formats.
    In principle, any expert knowledge, independent of its origin, that can be used to determine the coherence of explanations may be used to guide model attention. In practice, however, we recognize the significant difficulty of finding a suitable setup that would benefit from AGT. Partly due to the sparsity of meaningful domain priors that can be related to model explanations without requiring manual generation or annotation of such.
    
    Future research should aim to develop more principled and robust XAI methods tailored specifically for segmentation tasks. Establishing standardized evaluation protocols and benchmarks will be essential to assess explanation quality and model alignment across applications. 
    Ultimately, we aim for a more interactive and iterative approach to scientific machine learning, where explanations are not just used after the fact but become an integral part of training, evaluation, and scientific understanding.

\section{Methods} \label{sec:methods}
\begin{figure}[ht]
    \centering
    \includegraphics[width=.95\linewidth]{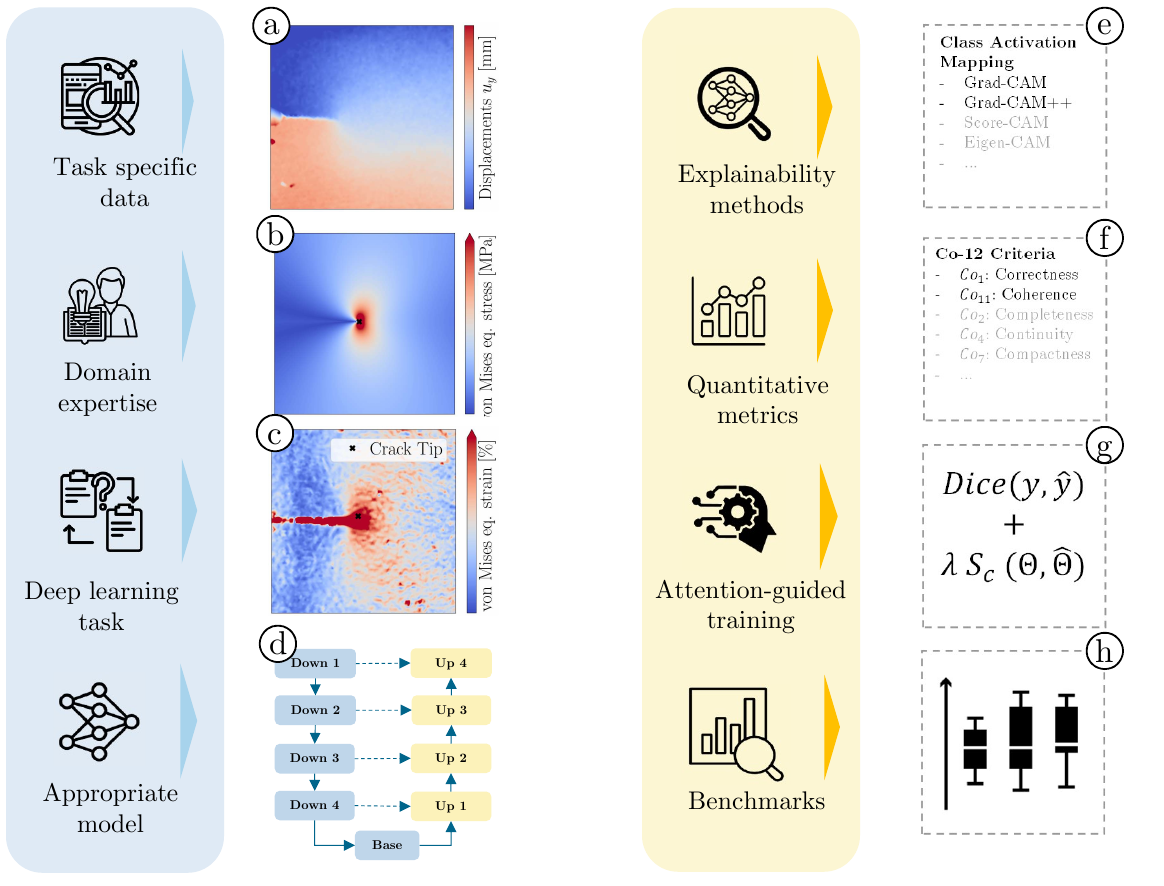}
    \caption{
    Visual summary of the task-specific AGT framework.\\
    \textbf{a)} Experimental DIC data (Section~\ref{sec:Data}).\\
    \textbf{b)} Crack tip specific domain knowledge via Williams series (Section~\ref{sec:AGT}).\\
    \textbf{c--d)} Crack tip segmentation task and U-Net architecture (Section~\ref{sec:MachineLearning}).\\
    \textbf{e--f)} CAM explanations and evaluation with Co12 criteria (Sections~\ref{sec:XAI}, \ref{sec:Co12}).\\
    \textbf{g)} Joint loss combining segmentation and explanation supervision (Section~\ref{sec:AGT}).\\
    \textbf{h)} Metrics to estimate model performance and robustness during generalization (Section~\ref{sec:MachineLearning})}
    \label{fig:visualabstract}
\end{figure}

To demonstrate attention-guided training (AGT), we applied it to crack tip segmentation in fatigue experiments. Figure~\ref{fig:visualabstract} shows the task-specific implementations of the general framework.

\subsection{Experimental data} \label{sec:Data}
We used previously published digital image correlation (DIC) data from three fatigue crack growth (FCG) experiments on AA2024-T3 middle-crack tension specimens~\cite{S160,S950,NextGenFatigue}.  
Displacement fields were interpolated on a $256\times256$ grid using CrackPy~\cite{Crackpy}, yielding two-channel inputs ($u_x,u_y$) for deep learning.
No new experiments were performed.

The dataset $S_{160,4.7}$ (166 samples at maximum force per side of the crack) was labeled using additionally obtained images of the polished sample surface. The data is split into training and validation based on sample side, where the crack growing to the right is used for training and the left for validation. The left side inputs are mirrored such that they match the right.
Two additional datasets, $S_{160,2.0}$ (280 samples) and $S_{950,1.6}$ (102 samples), provided out-of-distribution test cases with different geometries and loading conditions. Further details on the data and the generation of ground-truth labels are provided in our previous work~\cite{S160}.

\subsection{Machine learning} \label{sec:MachineLearning}
\subsubsection*{Task and architecture}
Crack tip detection was formulated as a binary semantic segmentation task~\cite{MelchExplainCrack}.
A U-Net~\cite{UNet} with LeakyReLU activations was used, mapping two-channel displacements to a single-channel crack-tip-probability map. 
The encoder consists of four convolutional blocks and a bottleneck, mirrored by a decoder with linear upsampling. A visual representation of this structure is depicted in Figure~\ref{fig:results_dltask}.

\subsubsection*{Data preparation}
Displacements were channel- and sample-wise normalized using the mean and standard deviation of each sample.
Annotated single-pixel crack tips were expanded to $5\times5$ (train) and $3\times3$ (validation) masks.
Augmentations included random crops (130--180 px), rotations ($\pm10^{\circ}$), and flips.
Validation and test data were not augmented.

\subsubsection*{Training}
Each training uses a randomly initialized U-Net with LeakyReLu activation functions \cite{MelchExplainCrack}.
Training used PyTorch with Adam ($5\times10^{-4}$, AMSGrad, batch size 16) and dropout ($p=0.3$) at the bottleneck.  
A two-stage strategy was applied: (i) 30 epochs of initial pretraining with segmentation loss only and (ii) 370 epochs with additional AGT supervision.  
Reference models were trained for 400 epochs without AGT.

\subsubsection*{Loss and metrics} \label{sec:Loss}
\textbf{Segmentation loss:} Dice loss addressed the strong class imbalance.

\textbf{Explanation loss:} Cosine similarity (CSI) measured alignment of model attention $\Phi$ with target attentions $\hat\Phi$.

\textbf{Reliability:} A prediction was valid if it contained exactly one connected region; reliability was the fraction of valid predictions:
\begin{equation}
    \text{Rel} = \frac{\# \text{ samples with exactly one segmentation}}{\# \text{ total samples}}.
\end{equation}

\subsection{Explainability} \label{sec:XAI}
Class activation mapping (CAM) was used to estimate model attention w.r.t. input features.  
Because segmentation outputs are spatial, we replaced the class score with the global average of output logits (following the approach from \cite{MelchExplainCrack} and \cite{SegGradCAM}).  
Explanations were generated from encoder and bottleneck layers, where features retain higher spatial fidelity.  
We evaluated gradient-based (Grad-CAM, Grad-CAM++, LayerCAM, etc.) and gradient-free methods (EigenCAM, Score-CAM, Ablation-CAM), normalizing the resulting explanations to $[0,1]$.  
For AGT, we selected Grad-CAM++ with encoder-layer aggregation.
We refer to Appendix \ref{app:XAI} for further details.

\subsection{Objective metrics for explainability} \label{sec:Co12}
To assess explanation quality, we used five criteria from Nauta et al.~\cite{Co12}:  
\textbf{Correctness} (here via incremental deletion), \textbf{Completeness} (here via incremental insertion), \textbf{Continuity} (here via SSIM across time steps), \textbf{Compactness} (here via the minimal input features required to recover a Dice coefficient of $\geq 0.8$), and \textbf{Coherence} (here via agreement with domain targets measured using \(S_c\)~\cite{Cosine}).
These metrics provided a quantitative basis for comparing CAM variants beyond visual inspection.
We refer to Appendix \ref{app:Co12} for further details on the equations and calibration details.

\subsection{Attention-guided training with domain knowledge} \label{sec:AGT}
Target attentions were derived from the near-tip stress field using the Williams series expansion.  
Two physical (binary, gradual Williams) and two unphysical baselines (binary and multi-gradual misleading) were considered as attention targets (see Appendix~\ref{app:target_attentions} for details).
AGT integrates prediction and explanation supervision into a joint loss:
\begin{equation}
    L_{\text{total}} = (1 - \text{Dice}(\mathbf{y}, \hat{\mathbf{y}})) + 
    \lambda \cdot \text{CSI}(\mathbf{\Phi}, \hat{\mathbf{\Phi}}),
\end{equation}
where $\mathbf{y}$ is the model output, $\hat{\mathbf{y}}$ the ground truth mask, $\Phi$ the Grad-CAM++ explanations, and $\hat{\Phi}$ the target attention maps.  
We used $\lambda=2$, chosen empirically.  
Training followed a two-stage scheme: initial segmentation pretraining, followed by AGT with attention supervision.
We refer to Appendix \ref{app:AGT} for further details on the training procedure and target attentions.

\section{Acknowledgements}
We acknowledge the financial support of the DLR-Directorate Aeronautics.

\section{Data availability statement}
All data are available on Zenodo (\url{https://doi.org/10.5281/zenodo.5740216}). The code for training and evaluation is archived on Zenodo (\url{https://doi.org/10.5281/zenodo.16902960}) and mirrored on Github (\url{https://github.com/dlr-wf/attention-guided-training}).

\section{Competing interests}
All authors are named inventors on the patent application DE102025113325.5 (pending), filed by the German Aerospace Center (DLR), which is related to the algorithm described in this work.

\section{Author contributions}
J.T. and D.M. conceived the attention-guided training framework. J.T. implemented the framework. E.B. provided the fracture mechanical domain knowledge. All Authors discussed, analyzed, and interpreted the results and wrote the manuscript.

\printbibliography

\appendix
\section{Explainability}
    \label{app:XAI}
    The model behavior can be approximately described using class activation mapping~\cite{CAM}.
    The CAM methods provide explanations in the form of spatial attention heatmaps indicating the importance of different input regions to the models prediction for a certain output class.
    These explanations are calculated as a linear combination of the spatial feature maps \(\mathbf{A}^k\) of a convolutional layer \(l\) with a set of relevance weights \(\omega^c_k\) associated with a class \(c\), yielding:
    \begin{equation}
        L^{c, l}_{\text{CAM}} = \sum_k \omega^c_k \mathbf{A}^k,
    \end{equation}
    which is postprocessed by applying ReLU and upsampled to the spatial dimensions of the input to obtain a class-specific attention map \(L^{c, l}_{\text{method}}\).
    \\
    The original CAM method is restricted to specific architectures, containing convolutional layers proceeded by a fully connected layer and applicable only to classification tasks \cite{GradCAM}. 
    Grad-CAM~\cite{GradCAM} was proposed to generalize this approach to more complex architectures.
    Many novel CAM techniques in the literature are slight variants of Grad-CAM.
    These typically operate on the final convolutional layer and assume a single class score as output. 
    Further changes have to be considered when applying these approaches to semantic segmentation~\cite{SegGradCAM}.
    \\
    Moreover, directly applying CAM to the final decoder layers of U-Net proved uninformative due to spatial sparsity and reduced feature variation in those layers.
    Approaches proposed by \cite{LayerCAM, MelchExplainCrack, SegGradCAM} consider earlier convolutional blocks—specifically, the encoder and bottleneck layers—where richer spatial structure is retained.
    To support explanation at multiple levels, we define block-wise outputs \texttt{Down1} through \texttt{Down4}, \texttt{Base}, and \texttt{Up1} through \texttt{Up4}, as described in Section \ref{sec:MachineLearning}.
    The multi-layer aggregation is defined here as the average over explanations calculated for multiple blocks:
    \begin{equation}
        L^{c,[l_1,\dots,l_n]}_{\text{method}} = \sum_{i=1}^{n} \beta_i L^{c,l_i}_{\text{method}}, \quad \beta_i = 1.
    \end{equation}
    
    \subsection*{Adapting CAM for segmentation.}
        Since semantic segmentation tasks produce spatial outputs, a scalar class score \(S^c\) is defined instead.
        \(S^c\) is calculated by aggregating logits over a subset of output pixels.
        Using the entire output mask retains the maximal amount of information:
        \begin{equation}
            \label{eq:GAPScore}
            S^c_{\text{GAP}} = \frac{1}{N} \sum_{i,j}^{m,n} f(\mathbf{X})^c_{ij}, \quad N = m \cdot n
        \end{equation}
        This score replaces the scalar output used in original CAM methods. Other score definitions were discussed in  \cite{SegGradCAM}.

    \subsection*{Gradient-based CAM methods.}
        Several gradient-based CAM methods were considered, all using the above score function \eqref{eq:GAPScore}:
        \begin{itemize}
            \item Grad-CAM \cite{GradCAM}
            \item Grad-CAM++ \cite{GradCAM++}
            \item Grad-CAM+ \cite{GradCAM+}
            \item LayerCAM \cite{LayerCAM}
            \item Elementwise-Grad-CAM \cite{GradCAMEW}
        \end{itemize}
    
    \subsection*{Gradient-free CAM methods.}
        Alternatively, gradient-free CAM methods were evaluated. In particular:
        \begin{itemize}
            \item EigenCAM \cite{EigenCAM}
            \item Score-CAM \cite{ScoreCAM}
            \item Ablation-CAM \cite{AblationCAM}
        \end{itemize}
        Eigen-CAM can be directly applied to intermediate feature maps without adaptation. 
        Score-CAM and Ablation-CAM typically rely on confidence drops in classification. 
        For segmentation, the degradation of a suitable prediction metric (e.g., logit aggregation, Dice score) is used as an alternative measure.
        All CAM outputs are normalized to \([0,1]\) and evaluated using objective criteria described in Appendix~\ref{app:Co12}.

\section{Objective metrics for explainability methods} \label{app:Co12}
    To ensure that model explanations are meaningful, trustworthy, and unbiased, a set of objective criteria is applied to evaluate and compare different CAM-based explainability methods.
    A variety of individual criteria to measure explanation quality can be found in literature, including perturbation-based evaluation measures~\cite{ROAD}, axiomatic or theoretical frameworks for explanation properties~\cite{Bach2015}, and human-aligned faithfulness metrics~\cite{yeh2019}. To our knowledge, no universal consensus has been reached so far.
    Following the Co-12 framework introduced by Nauta et al. \cite{Co12}, we adopt a principled subset of explanation quality criteria that are meaningful and operational for CAM-based explanations in crack-tip segmentation. While Co-12 provides a versatile and unifying taxonomy, the application domain and practical feasibility determines the emphasis on and selection of the different criteria.
    Thus, we restrict our evaluation to a task-appropriate subset of criteria, as explicitly encouraged to preserve objectivity and comparability. In our case, criteria that depend on class contrastivity, user interaction, uncertainty quantification, or subjective concept decomposition cannot be meaningfully defined or are not sufficiently relevant for the problem at hand. We therefore focus on correctness, completeness, continuity, compactness, and coherence, as these criteria capture complementary aspects of explanation quality while avoiding ill-posed or subjective measures, enabling a robust comparison and application of explainability methods for our crack tip segmentation task.
    \subsubsection*{Correctness (\(Co_{1}\))}
\begin{figure}[ht]
        \centering
        \includegraphics[width=\linewidth]{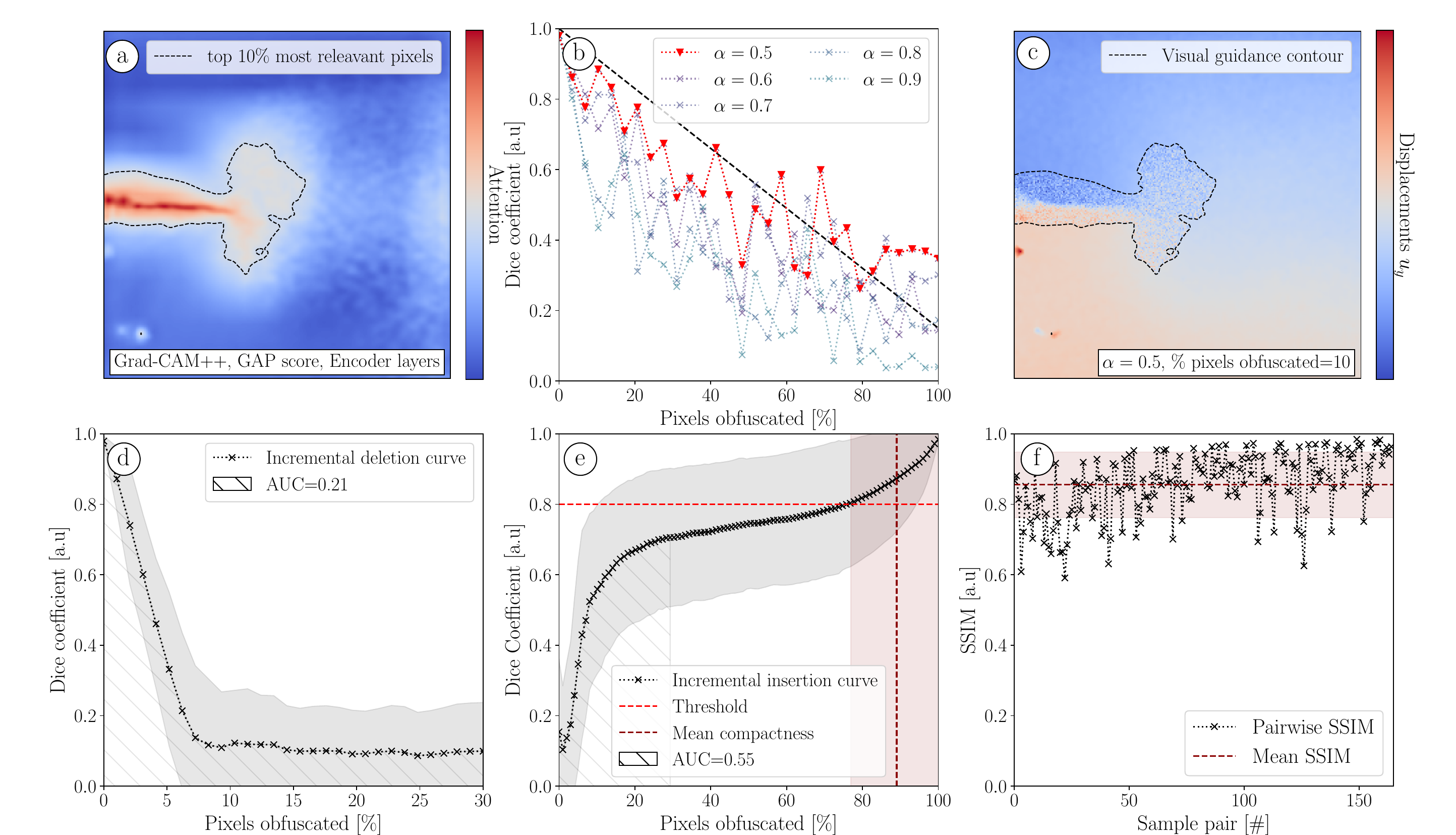}
        \caption{Quantitative evaluation of explanation quality using selected Co12 criteria. Mean and standard deviation are reported across all validation samples. 
        \textbf{a)} Example Grad-CAM++ attention heatmap computed from encoder layers using the GAP score. Pixels with the top 10\% of attention highlighted. 
        \textbf{b)} Calibration curves showing Dice degradation under random pixel obfuscation with varying noise scales $\alpha$, cf. Equation \eqref{eq:Co1}.
        \textbf{c)} Input displacement $u_y$ after Gaussian noise with $\alpha = 0.5$ is applied to the top 10\% most relevant pixels.
        \textbf{d)} Incremental deletion (correctness, \(Co_1\)) curve; steeper decline -- as measured by AUC -- indicates better correctness.  
        \textbf{e)} Incremental insertion (completeness, \(Co_2\)) and Compactness (\(Co_7\)) curve; AUC reflects completeness, while the threshold indicates how quickly the original performance is recovered. 
        \textbf{f)} continuity (\(Co_4\)) measured via pairwise SSIM across consecutive samples; higher values indicate explanation stability.}
        \label{fig:co12}
    \end{figure}

        Correctness (or faithfulness) measures how well an explanation reflects the true decision process of the model and thus is of utmost importance.
        To quantify correctness, incremental deletion was used, where input pixels are obfuscated in order of decreasing relevance as predicted by the corresponding explanation \cite{ROAD}.
        A fast degradation of the agreement between predictions made on obfuscated samples with respect to the original (as measured here by the Dice coefficient) is interpreted as evidence of explanation correctness.
        \\
        The Gaussian noise intensity is calibrated by randomized incremental deletion using:
        \begin{equation}
            \label{eq:gaussianObfuscate}
                (\tilde{\textbf{X}}_p)_{ij} = (\textbf{X})_{ij} + 
            \begin{cases}
                \alpha \mathcal{N}(\mu=0,\sigma=1) & \text{if} \quad  i,j \in M_p, \\
                0 & \text{otherwise,}
            \end{cases}
        \end{equation}
       where $\mathcal{N}$ denotes Gaussian noise, $p$ is the percentage of obfuscated pixels, $M_p$ denotes a set of $p$ percent of pixels, and $\tilde{\textbf{X}}_p$ is the obfuscated input with $p$ percent obfuscation. 
       The noise scale \(\alpha\) is calibrated per model, averaging over the results of multiple inputs. A well-calibrated obfuscation should cause an approximately linear decrease in prediction performance as the obfuscated set of pixels $M_p$ increases with $p$.
       An example of this approach is depicted in Figure \ref{fig:co12} b).
       \\
        The relevance estimates obtained through the CAM methods in combination with the calibrated obfuscation yield the incremental deletion technique.
        The inputs are deleted in order of relevance, where the area under the curve (AUC) for deletion pixel percentages \(p \in [0, P]\) defines the correctness score:
       \begin{equation}\label{eq:Co1}
            Co_1 = \int_0^P \text{Dice}(\sigma(f(\tilde{\mathbf{X}}_p)), \hat{\mathbf{y}} ) \, dp,
        \end{equation}
        where $\tilde{\mathbf{X}}_p$ denotes the obfuscated input and $\hat{\mathbf{y}}$ is the original binarized prediction. $Co_1$ is usually averaged over several inputs, e.g., a whole dataset, and several random Gaussian noises.
        Lower AUC indicates higher correctness of the explanation.
    \subsubsection*{Completeness (\(Co_{2}\))}
        Complementary to correctness, completeness measures how much relevant information is preserved in the most important regions of the input \cite{Co12}.
        It is typically assessed through incremental insertion, a process in which the most relevant pixels are gradually reintroduced into a fully obfuscated input to evaluate how well the model's predictions recover.
        The AUC of the resulting Dice curve (from \(p=0\) to \(p=P\)), similar to \eqref{eq:Co1}, defines the completeness score. Higher AUC indicates better completeness of the explanation.
    \subsubsection*{Continuity (\(Co_{4}\))}
        Continuity assesses whether small changes in the input lead to corresponding changes in the explanations.
        In fatigue crack growth (FCG) experiments, the DIC displacement data evolves smoothly over time. To evaluate continuity, predictions, and explanations are compared between consecutive samples in the sequence.
        Similarity of explanations is measured using the structural similarity index measure (SSIM) with the standard parameters introduced in \cite{ZhouWang2004}.  
        The continuity score is computed as the average SSIM over all consecutive samples in a dataset:
        \begin{equation}
            Co_4 = \frac{1}{N-1} \sum_{i=2}^{N} \text{SSIM}\left(\Phi(\mathbf{X}_{i-1})), \Phi(\mathbf{X}_i)\right)
        \end{equation}
    \subsubsection*{Compactness (\(Co_{7}\))}
        Compactness is evaluated by determining how quickly predictive accuracy recovers during the incremental insertion (completeness) curve.
        Specifically, the minimal percentage \(p^*\) of top-ranked pixels required to recover a Dice score \(\geq0.8\) is recorded.
        Methods with smaller \(p^*\) values are considered more compact.
    \subsubsection*{Coherence (\(Co_{12}\))}
        Coherence measures agreement between model explanations and (expert-derived) attention targets, such as the physical crack tip field.  
        Explanations \(\Phi\) and attention targets \(\hat{\Phi}\) are compared using SSIM and CSI.  
        This metric is not used to judge explanation quality in isolation but serves as the key component of our attention-guided training loss (see Appendix \ref{app:AGT}).
    \\
    \\
    These five metrics provide a multi-faceted, quantitative basis for comparing explanation methods beyond visual inspection or subjective plausibility.  
    All evaluations are conducted on the validation dataset.
    
\section{Attention-guided training}
    \label{app:AGT}
    To perform attention-guided training (AGT), domain-specific target attentions are added to the training and validation datasets as supervision signals.
    For fatigue crack growth in samples with a small plastic zone w.r.t. the crack length, the Williams series was chosen as a well-established and robust analytical representation of the stress field near the crack tip.
    A per-sample parameterization of Equation \ref{eq:williams_stress} was omitted for simplicity, using instead a fixed, representative configuration.
    
    The parameters were obtained using a least-squares fit method, available in the CrackPy library, yielding:
    \begin{itemize}
        \item \(A_1 = \frac{K_I}{\sqrt{2 * \pi}} = \frac{23.71 * \sqrt{1000} }{\sqrt{2 * \pi}}\) MPa\(\sqrt{\text{mm}}\)
        \item \(A_2 = \frac{T}{4} = \frac{-7.22 }{4}\) MPa 
        \item \(A_3 = -2.87 \) \(\frac{\text{MPa}}{\sqrt{mm}}\)
        \item \(A_4 = 0.03 \) \(\frac{\text{MPa}}{mm^{-1/4}}\)
        \item \(A_n = 0\; \forall \; n>4\)
        \item \(B_n = 0\; \forall \; n\)
    \end{itemize}
    
    The proposed attention-guided training (AGT) framework integrates the components described above into a unified approach that explicitly steers model attention toward domain-specific priors during training.
    Model explanations, generated using CAM techniques, are evaluated using four objective criteria -- correctness, compactness, continuity, completeness -- and are aligned with target explanations derived from fracture mechanics. This alignment is enforced through a coherence criterion incorporated into the overall loss function.
    \\
    The approach follows a two-stage training procedure:
    \begin{enumerate}
        \item An initial pretraining phase using only prediction loss.
        \item A joint training phase using both prediction and explanation loss.
    \end{enumerate}
    \subsection*{Model initialization and explanation selection}
        The U-Net model \cite{UNet} is initialized using random weights. 
        Initial training is required to reach a state of salient explanations; this phase is empirically set to 30 epochs.
        Explanations are then generated using the selected CAM methods mentioned in Appendix \ref{app:XAI}.
        Variations of targeted layers, methods, and score functions were considered.
        Preliminary qualitative evaluations and discussions led to a focused exploration of the encoder branch (Blocks \texttt{Down1} - \texttt{Down4} and \texttt{Base}) in combination with the GAP score function.
        This choice is supported by several arguments: early encoder layers preserve diverse, high-fidelity features; deeper layers (e.g., the bottleneck) capture more abstract representations; and GAP considers the entire input signal when computing explanation scores. 
        Finally, Grad-CAM++ was chosen for AGT using the metrics found in Figure \ref{fig:co12eval}.
        
    \subsection*{Target attentions} \label{app:target_attentions}
        Target attention maps \(\hat{\Phi}\) were computed from the von Mises stress field using the Williams series expansion (Equation~\ref{eq:williams_stress}) and manual crack tip annotations. 
        For comparison, two physical (domain-informed) and two unphysical attention target strategies were considered.
        The variations originating from the physical crack tip field are:
        \begin{itemize}
            \item \textit{Binary Williams (BW)}: The continuous field is binarized using an empirical threshold of 162 MPa. Regions above this threshold were set to 1, others to 0.
            \item \textit{Gradual Williams (GW)}:  The continuous field is obtained by clipping the Williams stress field to 162 MPa and below 75 MPa. Values below the threshold were set to 0, and values above to 1. The intermediate region was re-scaled, resulting values in the range $[0, 1]$, yielding a continuous, non-binary attention map with gradually fading intensity.
        \end{itemize}
        For comparison, additional unphysical attention targets were considered:
        \begin{itemize}
            \item \textit{Binary misleading (BM)}: The target attention is set to 1 within a 76-pixel square located at the bottom-right corner of the domain and 0 elsewhere.
            \item \textit{Multi-gradual misleading (MGM)}: The target attention is set to 1 at the top- and bottom-right corners and gradually decays within a radius of 70 pixels, using a radius-dependent exponential decay \(e^{-0.013 \cdot r}\).
        \end{itemize}
        The empirical parameters chosen here were determined by balancing total relevant pixels and attention with (ir)relevant areas. All attention maps were clipped and normalized to \([0,1]\) for comparability.
        
    \subsection*{Training procedure}
        Subsequently, attention supervision was activated via a joint loss:
        \begin{equation*}
            L_{\text{total}} = (1 - \text{Dice}(\mathbf{y}, \hat{\mathbf{y}})) + \lambda \cdot \text{CSI}(\mathbf{\Phi}, \hat{\mathbf{\Phi}}),
        \end{equation*}
        where \(y\) is the model output, \(\hat{\mathbf{y}}\) the binary segmentation label, \(\mathbf{\Phi}\) the Grad-CAM++ attention heatmaps, and \(\hat{\mathbf{\Phi}}\) the target attention heatmaps.  
        The cosine similarity (CSI) was chosen for its scale invariance and stable gradients.  
        A weighting factor \(\lambda = 2\) was used to balance the two loss terms, selected empirically to ensure that domain supervision did not dominate training.

\section{Statistical tests}
    \label{app:stats}
    To assess statistically significant differences between attention guidance strategies, we employed the Mann-Whitney-U (MWU) test.
    The MWU test is a non-parametric alternative to the two-sample $t$-test and does not assume normality, making it suitable for the small sample size of $n=10$, independent runs per strategy, and potentially skewed performance metrics considered in this study.
    All tests were conducted using the implementation contained in the Python package \textit{SciPy}.
    
    Each training run constitutes one independent sample.
    Runs were grouped according to their attention targets:
    \textit{Binary Williams (BW)}, \textit{Gradual Williams (GW)}, \textit{Binary Misleading (BM)}, \textit{Multi-Gradual Misleading (MGM)},
    and a \textit{Reference} configuration without attention guidance.
    For high-level analysis, strategies were further pooled into
    \textit{physical} (BW + GW) and \textit{misleading} (BM + MGM) groups.
    
    Two families of directional hypotheses were evaluated at a significance level of $\alpha=0.05$.
    First, three pre-specified overview comparisons were performed to assess the general effect of attention guidance:
    (i) Physical vs.\ Reference,
    (ii) Misleading vs.\ Reference, and
    (iii) Physical vs.\ Misleading.
    Second, four follow-up comparisons were conducted to assess the performance of the best-performing attention strategy (BW) relative to all remaining strategies.
    Directional hypotheses reflect the expected improvement direction
    (lower is better for validation loss and correctness AUC, higher is better for reliability).

    \begin{table}
    \centering
    \caption{
    Directional Mann-Whitney-U test results comparing attention guidance strategies.
    Each row reports the $p$-value for the stated directional hypothesis.
    Statistically significant results ($p < 0.05$) are highlighted in bold.
    }
    \label{tab:mwu_results}
    \begin{tabular}{l | c | c }
    \hline
    Metric & Hypothesis & $p$-value \\
    \hline
     \multirow{7}*{Validation loss} & \textbf{Physical} $<$ \textbf{Reference} & \textbf{0.0001} \\
     & \textbf{Physical} $<$ \textbf{Misleading} & \textbf{0.0} \\
     & Misleading $<$ Reference & 0.105 \\
     & \textbf{BW} $<$ \textbf{R} & \textbf{0.0018} \\
     & \textbf{BW} $<$ \textbf{BM} & \textbf{0.027} \\
     & \textbf{BW} $<$ \textbf{MGM} & \textbf{0.0009} \\
     & BW $<$ GW & 0.7146 \\
    \hline
    \multirow{7}*{Reliability $S_{160,4.7}$} & Physical $>$ Reference & 0.1586 \\
     & Physical $>$ Misleading & 0.124 \\
     & Misleading $>$ Reference & 0.4821 \\
     & BW $>$ R & 0.121 \\
     & \textbf{BW} $>$ \textbf{BM} & \textbf{0.0373} \\
     & BW $>$ MGM & 0.3758 \\
     & BW $>$ GW & 0.3909 \\
    \hline
    \multirow{7}*{Reliability $S_{160,2.0}$} & Physical $>$ Reference & 0.1778 \\
     &  Physical $>$ Misleading & 0.3149 \\
     &  Misleading $>$ Reference & 0.3038 \\
     &  BW $>$ R & 0.5616 \\
     &  BW $>$ BM & 0.7473 \\
     &  BW $>$ MGM & 0.8004 \\
     &  BW $>$ GW & 0.9836 \\
    \hline
    \multirow{7}*{Reliability $S_{950,1.6}$} & Physical $>$ Reference & 0.3789 \\
     & Physical $>$ Misleading & 0.2161 \\
     & Misleading $>$ Reference & 0.7015 \\
     & \textbf{BW} $>$ \textbf{R} & \textbf{0.0154} \\
     & \textbf{BW} $>$ \textbf{BM} & \textbf{0.0069} \\
     & \textbf{BW} $>$ \textbf{MGM} & \textbf{0.0045} \\
     & \textbf{BW} $>$ \textbf{GW} & \textbf{0.002} \\
    \hline
    \multirow{7}*{AGT correctness (smaller is better)} & \textbf{Physical} $<$ \textbf{Reference} & \textbf{0.0003} \\
     & \textbf{Physical} $<$ \textbf{Misleading} & \textbf{0.0} \\
     & Misleading $<$ Reference & 0.2476 \\
     & \textbf{BW} $<$ \textbf{R} & \textbf{0.0036} \\
     & \textbf{BW} $<$ \textbf{BM} & \textbf{0.0036} \\
     & \textbf{BW} $<$ \textbf{MGM} & \textbf{0.0129} \\
     & BW $<$ GW & 0.8276 \\
    \hline
    \end{tabular}
    \end{table}

    Overall, physical attention guidance yields statistically significant improvements in validation loss across both grouped and individual strategy comparisons.
    Reliability differences are largely non-significant for in-distribution and mildly out-of-distribution datasets, which exhibit near-saturated performance across all strategies.
    Statistically meaningful reliability differences emerge only for the far out-of-distribution dataset $S_{950,1.6}$, where the BW target consistently outperforms all baselines.
    These findings indicate that physical attention guidance improves optimization behavior and robustness under distribution shift, with the BW attention targets achieving the strongest and most consistent performance gains.

\end{document}